\documentclass[iop]{emulateapj}
\usepackage{apjfonts}
\usepackage{graphicx}
\usepackage{amsmath}
\usepackage{epstopdf}

\slugcomment{Prepared for ApJ}

\begin{document}

\title{Expectation on Observation of Supernova Remnants with the LHAASO Project}

\author{Ye Liu\altaffilmark{1}, Zhen Cao\altaffilmark{2}, Songzhan Chen\altaffilmark{2}, Yang Chen\altaffilmark{3}, Shuwang Cui\altaffilmark{4}, Huihai He\altaffilmark{2}, Xingtao Huang\altaffilmark{1;*}, Xinhua Ma\altaffilmark{2;$\dagger$}, Qiang Yuan\altaffilmark{5}}
\author{ Xiao Zhang\altaffilmark{3} \centerline {On behalf of the LHAASO Collaboration}}


\altaffiltext{*}{Corresponding author: huangxt@sdu.edu.cn}
\altaffiltext{$\dagger$}{Corresponding author: maxh@ihep.ac.cn}
\altaffiltext{1}{School of Physics $\&$ Key Laboratory of Particle Physics and Particle Irradiation(MOE), Shandong University, Jinan 250100, China}
\altaffiltext{2}{Institute of High Energy Physics, Chinese Academy of Sciences, Beijing 100049, China}
\altaffiltext{3}{Department of Astronomy, Nanjing University, Nanjing 210093, China}
\altaffiltext{4}{The College of Physics Science and Information Engineering, Hebei Normal University, Shijiazhuang, 050016, China}
\altaffiltext{5}{Department of Astronomy, University of Massachusetts, 710 North Pleasant St., Amherst, MA, 01003, U.S.A}

\begin{abstract}
Supernova remnants (SNRs) are believed to be the most important
acceleration sites for cosmic rays (CRs) below $\sim10^{15}$ eV in the
Galaxy. High energy photons, either directly from the shocks of the SNRs
or indirectly from the interaction between SNRs and the nearby clouds, are crucial probes
for the CR acceleration. Big progresses on observations of SNRs have been
achieved by space- and ground-based $\gamma$-ray facilities. However, whether $\gamma$-rays come from accelerated
hadrons or not, as well as their connection with the CRs observed at
Earth, remains in debate. Large High Altitude Air Shower Observatory
(LHAASO), the next generation experiment, is designed to survey the northern part of the
very high energy $\gamma$-ray sky from $\sim 0.3$ TeV to PeV  with the sensitivity of $\lesssim1\%$ of the Crab nebula flux. In this paper,
we indicate that LHAASO will be dedicated to enlarging the $\gamma$-ray SNR
samples and improving the spectral and  morphological measurements.
These measurements, especially at energies above 30 TeV, will be important for us to finally understand the CR acceleration in SNRs.

\end{abstract}
\keywords{acceleration of particles - cosmic rays - gamma rays}
\section{introduction}

Cosmic rays (CRs), discovered more than one century ago, represent the
most energetic, and, to some degree, fundamental aspect of the Universe.
Nuclei constitute dominantly ($\sim99\%$) CRs, and electrons contribute
the rest $\sim1\%$. The energy spectrum of CRs observed at Earth is well
described by a single power-law up to energies of a few PeV, above which
the spectrum steepens to form the so-called ``knee'' \citep{Horandel2003,Blumer2009}.
It is general believed that CRs below the ``knee'' originate from the
Galaxy, due to a simple argument that the Galactic magnetic field would
effectively confine such particles. However, their acceleration sites and
the acceleration mechanism remain open questions.

The deflection of charged CRs in the Galactic magnetic field makes it
very difficult to identify the sources of CRs. High energy $\gamma$-rays,
produced by either the collision between CR hadrons and the ambient
medium or the interaction between CR leptons and the interstellar radiation
field and/or the medium, are thus more powerful to probe the sources of CRs.
With quick development of the detection technology, great progresses
have been achieved in high energy $\gamma$-ray astronomy during the past
two decades. The satellite-borne instrument Fermi Large Area Telescope
(Fermi-LAT), launched in 2008, discovered over 3000 sources in GeV energies
\citep{Acero2015a}. In TeV energies, more than 160 $\gamma$-ray sources$^1$
have been firmly detected by the ground-based detectors,
including Imaging Atmospheric Cherenkov Telescopes (IACTs) such as H.E.S.S.,
MAGIC, and VERITAS, and extensive air shower (EAS) arrays such as
Tibet-AS$\gamma$, Milagro, and ARGO-YBJ. Most of these $\gamma$-ray sources
are likely to be leptonic, including pulsars, pulsar wind nebulae (PWN),
binary systems and so on \citep[e.g.,][]{Aharonian2006,Chen2013,Bartoli2015}.
Clear evidence of the sources of hadronic CRs 
is observationally rare.

It is widely believed that supernova remnants (SNRs) could efficiently
accelerate particles at the shock front where the expanding supernova
ejecta encounter the surrounding medium \citep{Bell1978a,Bell1978b,Drury1994}.
The well-established diffusive shock acceleration (DSA) theory
\citep{Drury1983,Blandford1987} predicts a power-law spectrum of the
accelerated particles with index of $\sim2$ \citep{Holder2012,Schure2012},
well consistent with the radio observations of SNRs$^2$ ,
as well as the locally observed CR spectrum accounting for the propagation
effect. Moreover, the importance of magnetic field amplification by
the accelerated particles themselves has been increasingly recognized,
which is expected to play a particularly important role in explaining
the CRs around the knee region ($\sim$PeV) \citep{Bell2004}.

\let\thefootnote\relax\footnote{1 http://tevcat.uchicago.edu/}
\let\thefootnote\relax\footnote{2 http://www.mrao.cam.ac.uk/surveys/snrs/}
Based on the multi-wavelength observations of SNRs, a class of SNRs have
been suggested to be efficient electron accelerators, e.g., RX J1713.7-3946
\citep{Ellison2010,Abdo2011,Li2011,Yuan2011}, RX J0852-4622
\citep{Tanaka2011}, RCW 86 \citep{Yuan2014}, HESS J1731-347
\citep{Yang2014,Acero2015b}, and SN 1006 \citep{Acero2010,Araya2012,Acero2015b}.
On the other hand, Fermi-LAT collaboration  reported  the detection
of the characteristic pion-decay bump in the GeV $\gamma$-ray spectra of
two ancient SNRs, IC~443 and W44, which are interacting with molecular
clouds \citep{Ackermann2013}. An earlier detection of hard sub-GeV spectrum
of W44 by AGILE was also explained to be hadronic origin \citep{Giuliani2011}.
These observations might give direct evidence for hadronic CR acceleration
in SNRs \citep{Ackermann2013}. However, the role of electron bremsstrahlung
in this energy range is still unclear, due to the poorly constrained sub-GeV
electron spectrum.

The observations of very high energy (e.g., $>100$ TeV) $\gamma$-ray
emission from SNRs will provide a complimentary test of the leptonic/hadronic
origin of the $\gamma$-rays. It is expected that electrons will be difficult
to be accelerated sufficiently to high energies due to the strong synchrotron
and/or inverse Compton cooling. A rough estimate of the synchrotron
cooling gives the cooling energy, $E_c\sim10\,(B/\mu{\rm G})^{-2}
(T/10^3\,yr)^{-1}$ PeV, above which electrons will cool down. For a typical
magnetic field strength of tens of $\mu$G and an age of $10^3$ yr,
$E_c\lesssim100$ TeV. The acceleration of high energy electrons will be
even more difficult for old SNRs. Furthermore, the inverse Compton radiation
efficiency of very high energy electrons will be suppressed due to the
Klein-Nishina effect, making the contribution to very high energy
$\gamma$-rays more difficult.
It is thus expected that $\gamma$-ray emission
above $100$ TeV from SNRs (especially the old ones), either from shock-crushed dense clouds \citep[e.g.][]{Blandford1982, Uchiyama2010} or from the adjacent MCs illuminated by escaping
protons \citep{Li2010,Li2012}, will strongly support
the nuclei acceleration.
Besides, the $\gamma$-ray observation above 100 TeV is also expected to give a constraint on the cutoff energy of shock-accelerated proton, which can help to make a distinction between the theoretical models \citep{Zhang2016}.
These become the major motivations that
we propose the next generation, km$^2$ scale CR/$\gamma$-ray observatory:
the Large High Altitude Air Shower Observatory \citep[LHAASO;][]{Cao2010,He2015}
. With the large effective area ($\sim$km$^2$) and large
field-of-view ($\sim$2 sr), as well as the high CR rejection power
($\sim$1\%), LHAASO is dedicated to surveying the northern very high
energy $\gamma$-ray sky with sensitivity of $\sim1\%$ of the Crab nebula
flux. LHAASO is expected to not only discover many new sources, but also
improve the spectral and morphological measurements of known sources
to the highest achievable energies \citep{Cui2014,Zhao2016}.

In this work we study the perspective of observing very high energy
$\gamma$-ray emission from SNRs with LHAASO. We intend to understand
the physical potential of LHAASO on the long-standing problem of
the origin of hadronic CRs. The rest of this paper is organized as follows.
We briefly introduce the design of LHAASO detectors and their performances
on $\gamma$-ray detections in Sec. 2. The expected performance of SNR
observations with LHAASO based on the current GeV and/or TeV $\gamma$-ray
observations is presented in Sec. 3. We will pay special attention on the
spectral measurements to $>30$ TeV energies. We conclude our work with
some discussions in Sec. 4.

\section{the LHAASO project and its performance}

LHAASO (100.01$^{\circ}$E, 29.35$^{\circ}$N) is a recently approved, new
generation EAS experiment which will be built at 4410 m above the sea level near the Daocheng village,
in Sichuan province, China. LHAASO is a hybrid extensive air shower (EAS) array covering an area of 1 km$^2$ (KM2A),
water Cherenkov detector array (WCDA),  and wide field-of-view imaging
Cherenkov telescope array (WFCTA). The major scientific goals of LHAASO are: 1) precisely measuring
the spectra of $\gamma$-ray sources at high energy range for studying the acceleration and propagation of CRs,
2) deeply surveying the very high energy $\gamma$-ray sky (Dec. from $-10^{\circ}$ to $70^{\circ}$) for exploring the high energy radiation mechanism, and 3) effectively
searching for dark matter and the new physics \citep{Cao2010,He2015}.
The setup of LHAASO (see Fig. 1) and its performance are as follows.

\begin{figure}
\includegraphics[width=3.5in,height=2.3in]{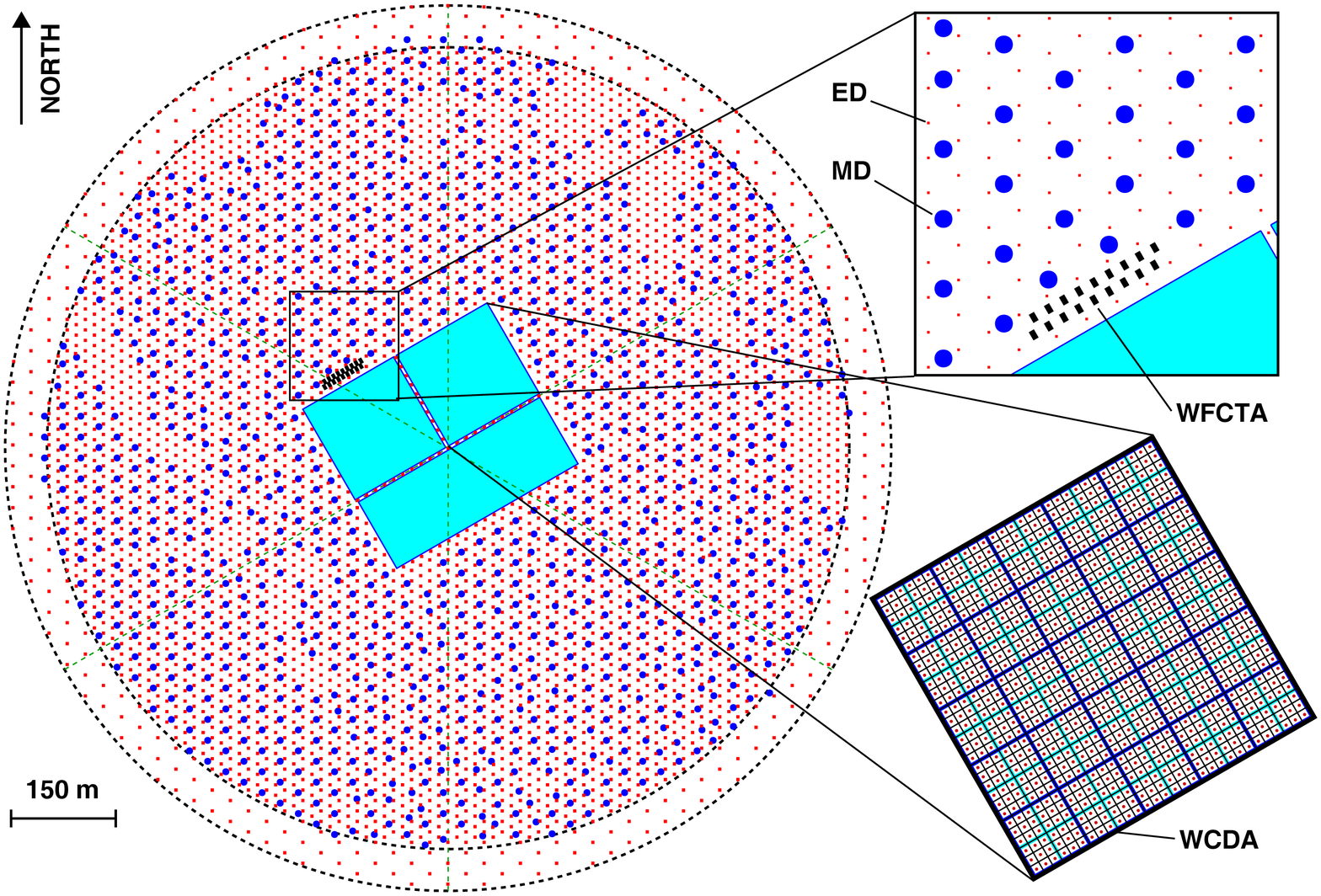}
\caption{Schematic plot of LHAASO detectors.}
\label{fig1}
\vspace*{0.5cm}
\end{figure}

$\bullet$ KM2A, with an effective area of 1 km$^2$,
is composed of 5195 scintillator electron detectors (EDs) with
1 m$^2$ each and a spacing of 15 m, and 1171 muon detectors (MDs) with 36
m$^2$ each and a spacing of 30 m.
The detector performance are discussed in (\cite{Cui2014,He2015}).
At 10 TeV, the effective area of KM2A can reach about 0.3 km$^2$, the angular resolution is about 0.86$^{\circ}$,
and the energy resolution for $\gamma$-rays is about 42$\%$. The
corresponding values are 0.8 km$^2$, 0.5$^{\circ}$, 33$\%$ at 30 TeV, and 0.9 km$^2$, 0.3$^{\circ}$, 20$\%$ at 100 TeV respectively.
With the large area of MD array, KM2A will reject the hadronic shower background at a level of $10^{-4}$ at 50 TeV and even $10^{-5}$ at higher energies,
so that $\gamma$-rays samples can reach background free above 100 TeV. The highest sensitivity of KM2A is $\sim1\%$ of the Crab
nebula flux in the energy range of $50-100$ TeV for one year observation.

$\bullet$ WCDA, with an effective area of 78,000 m$^{2}$ and 3000 units, will be built at the center of KM2A.
The size of one unit is about 5 m $\times$ 5 m, and the effective water
depth is about 4.4 m. Each unit is separated by plastic curtains vertically
hung in water, and contains a photomultiplier tube anchored at the center of the cell
bottom.
The detector performance are discussed in (\cite{Yao2009,Yao2011}).
 At 0.5 TeV, the effective area can reach 3,000 $m^2$, the angular resolution is 0.6$^{\circ}$ and the energy resolution for $\gamma$-rays is 95$\%$. The corresponding values are 10,000 m$^2$, 0.4$^{\circ}$, 90$\%$ at 1 TeV, and 50,000 m$^2$, 0.2$^{\circ}$, 60$\%$ at 10 TeV.
The highest sensitivity is $\sim1\%$ of the Crab nebula flux at the energy around 3
TeV.

$\bullet$ WFCTA, made up of 12 telescopes, will be
built near WCDA. These detectors are to measure the energy spectral of different composition of CRs
up to the second knee, less relevant to the $\gamma$-ray detection as
discussed in this work.

 \begin{figure}
 \centering
\includegraphics[width=3.5in,height=2.3in]{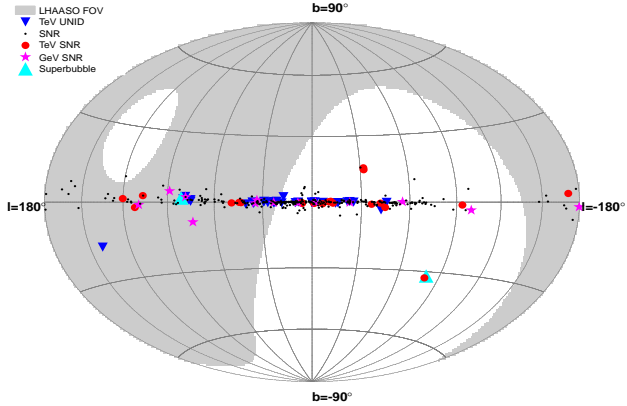}
  \caption{Locations of SNRs and unidentified TeV $\gamma$-ray sources in
Galactic coordinates, compared with the field-of-view of LHAASO (grey
region). Black dots represent SNRs from Green$^1$, red filled
circles and magenta stars show TeV and GeV $\gamma$-ray SNRs$^1$
\citep{Acero2015a}, blue triangles represent the unidentified TeV
$\gamma$-ray sources, and cyan triangles represent two superbubbles
which were detected in TeV $\gamma$-ray bands. (A color version of this figure is available in the online journal.)\label{fig2}}
  \label{fig2}
\vspace*{0.5cm}
 \end{figure}

\section{Expectation on observation of SNRs with LHAASO }

\subsection{Simulation of observing a $\gamma$-ray source}
To estimate LHAASO observation for a $\gamma$-ray source, both $\gamma$-ray and cosmic ray background samples are produced  by Corsika 6.600.
The detector response of WCDA is simulated based on GEANT4 program developed by the Milagro collaboration \citep{Yao2009}. For KM2A,   a fast simulation presented in \citep{Cui2014} is adopted. The simulation events are sampled in energy range from 100 GeV to 10 PeV with the zenith angle less than 45$^{\circ}$.

In order to obtain the signals (N$_{\gamma}$) and background (N$_{bg}$) within a specific space angle center on the source, we have traced the source   by means of a complete transit, i.e., 24 hours of observation. We only adopt the proton to simulate the cosmic ray background, while we  adjusted the flux to the all particles flux \citep{Ivanenko1988}.
We use Eq.(17) of Li and Ma \citep{Li1983} to  calculate the significance.

\subsection{Detectability of SNRs and SNR candidates}
According to Dave Green's Galactic SNR catalogue$^1$ \footnote{$^1$ http://www.mrao.cam.ac.uk/surveys/snrs/}, 294 SNRs have been detected up to now. Most of these SNRs are detected in low energy bands.
In GeV energies,  the Fermi-LAT collaboration reported their first  SNR catalog
based on three year's survey data, in which 12 firm identifications
and 11 possible associations with SNRs were found \citep{Acero2015a}. In TeV
energies, there have been 23 SNRs or SNR candidates detected up to now,
10 of which are also GeV $\gamma$-ray emitters$^2$ \footnote{$^2$ http://tevcat.uchicago.edu/}.
Furthermore, there are 34 unidentified TeV $\gamma$-ray sources which do
not have clear counterparts in other wavelengths. Different from the Fermi
unidentified sources which are expected to be dominately constituted
by active galactic nuclei \citep {Mao2013}, most of the unidentified
TeV sources are located in the Galactic plane (see Fig. 2) and could
be potential SNRs. Fig. 2 illustrates the locations of those sources
(symbol) and their visibility by LHAASO (shaded region). In total,
92 out of 294 SNRs in Green Catalog,  6 GeV SNRs or SNR candidates, 2 TeV SNRs and 6 GeV-TeV SNRs are in the field of view of LHAASO. Besides,
 17 TeV unidentified sources locate in the field of view of LHAASO.

Based on the current GeV/TeV observations of SNRs and SNR candidates,
we estimate their detectability by LHAASO. The $\gamma$-ray spectrum
of each source is assumed to be an exponentially cutoff power-law form
\begin{equation}
\frac{dN}{dE} = J_0(E/{\rm TeV})^{-\alpha}\exp(-E/E_{\rm cut})
\end{equation}

where $J_0$ is the flux at 1 TeV, $\alpha$ is the source spectral index, and $E_{\rm cut}$ is the cutoff energy. We fit the observational data
to find the spectral parameters for each source. For the GeV only sources
and some TeV sources, whose cutoff energies can not be well determined,
we assume $E_{\rm cut}=30$ and 100 TeV, respectively.

We simulate the observation of each source and calculate its expected
significance for 5 years sky survey of LHAASO.
For GeV-TeV and TeV SNRs, the expected significance above 10 TeV under cutoff energy 30 TeV and 100 TeV are presented in Table 1.
6 SNRs can been detected by LHAASO with significance greater than 5 $\sigma$ if the cutoff energy is 100 TeV. The significance will decrease if the cutoff energy is lower, while, there still are 5 SNRs can been detected by LHAASO if the cutoff energy is 30 TeV.
This is the key to confirm whether the SNR is hadronic origin or not, shows that the observation level of LHAASO is sensitive to give an judgement on the acceleration mechanism above 30 TeV.

Morever, it has been found that some SNRs could emitted TeV $\gamma$-rays while in GeV energy band there was no observation results, like G106.3+2.7 and HESS J1912+101. G106.3+2.7 was first observed by DRAO at radio energy range \citep{Joncas1990}. In 2000, Pineault $\&$ Joncas confirmed the object as a SNR, with an estimated age of 1.3 Myr and distance of 12 kpc \citep{Pineault2000}. The pulsar PSR J2229+6114 is located at the northern edge of the remnant's head and it is associated with boomerang-shaped radio and X-ray emitting wind nebula. At GeV energy band, the EGRET source 3EG J2227+6122 is compatible with the pulsar position, as well as the main bulk of the radio remnant \citep{Hartman1999}. And at TeV energy band, VERITAS reported the total flux from the SNR G106.3+2.7 above 1 TeV is about $\sim$5$\%$ of the Crab Nebular in 2009 \citep{Acciari2009b}. HESS J1912+101 is plausibly associated with the PSR J1913+1011, which is detected by H.E.S.S. experiment. The integral flux between 1-10 TeV is 10$\%$ of the Crab Nebula and the measured energy spectrum can be described by a power-law with a photon index $\sim$ 2.7. From the current observation on these two TeV SNRs, we can conclude that LHAASO might discover a number of SNRs compared to conservative predictions based on the current SNR catalogs.

\begin{table*}
\centering
\caption{Table of a selection of known GeV-TeV and TeV SNRs shown with the expected significance above 10 TeV using five years Monte Carlo simulation datas of LHAASO, assuming the sources are under the specific hypotheses for the energy spectrum. 
(``-" means the significance is less than 5$\sigma$.) }
\label{tab-1}
\vspace{1mm}
\begin{tabular}{ccccccccc}
  \hline\hline
  Name  & Classification & R.A.  & Dec. & $\alpha$ & $J_0$ & $\sigma$ &  $\sigma'$ & Ref.\\
        &                &       &      &          & (TeV$^{-1}$cm$^{-2}$s$^{-1}$) & ($E_{\rm cut}=100$ TeV) & ($E_{\rm cut}=30$ TeV) & \\

  \hline
  Tycho(a)     &GeV-TeV   & 00$^h$25$^m$18$^s$     &  +64$^{\circ}$09$^{'}$    & 2.92$\pm$0.46 & 2.2$\times$10$^{-13}$   &  -    &  -      & 1  \\

  \hline
  Tycho(b)     &GeV-TeV   & 00$^h$25$^m$18$^s$     &  +64$^{\circ}$09$^{'}$    & 1.95$\pm$0.51 & 1.70$\times$10$^{-13}$  & 11.62 &  5.20   & 2  \\


  \hline
  IC~443       &GeV-TeV   & 06$^h$17$^m$00$^s$     &  +22$^{\circ}$30$^{'}$    & 2.99$\pm$0.38 & 8.38$\times$10$^{-13}$  &  5.46 &  -     & 3  \\

  \hline
  W49B         &GeV-TeV   & 19$^h$11$^m$08$^s$     &  +09$^{\circ}$06$^{'}$    & 3.1$\pm$0.3   & 2.3$\times$10$^{-13}$   &  -    &  -      & 4  \\

  \hline
  HESS J1912-101& TeV     & 19$^h$12$^m$49$^s$     &  +10$^{\circ}$09$^{'}$    & 2.7$\pm$0.2   & 3.5$\times$10$^{-12}$   & 59.63 & 28.02   & 5  \\

  \hline
  W51C         &GeV-TeV   & 19$^h$23$^m$50$^s$     &  +14$^{\circ}$06$^{'}$    & 2.58$\pm$0.07 & 9.7$\times$10$^{-13}$   & 31.44 & 14.87   & 6  \\

  \hline
  G106.3+2.7   &TeV       & 22$^h$27$^m$59$^s$     &  +60$^{\circ}$52$^{'}$    & 2.29$\pm$0.33 & 1.42$\times$10$^{-12}$  &  57.43  & 21.10 & 7  \\

  \hline
  Cassiopeia~A &GeV-TeV   & $ 23$$^h$23$^m$26$^s$  &  +58$^{\circ}$48$^{'}$    & 2.3$\pm$0.2   & 7.3$\times$10$^{-13}$   &  26.51  & 10.20 & 8  \\

  \hline\hline
 \end{tabular}
\vspace*{0.5cm}

{ Notes: Columns from left to right are: source name, classification,
R.A., Declination, spectral index, flux normalization, cutoff energy,
expected significance by LHAASO (for $E_{\rm cut}=100$ TeV, if the cutoff
energy has not been measured), expected significance by LHAASO for
$E_{\rm cut}=30$ TeV, and the references of the measurements.\\

References: {(1) \citet{Park2015}; (2) \citet{Acciari2011}; (3) \citet{Acciari2009a}; (4) \citet{Francois2011}; (5) \citet{Aharonian2008b}; (6) \citet{Aleksic2012}; (7) \citet{Acciari2009b}; (8) \citet{Albert2007b};

 }}

 \end{table*}

\begin{table*}
\centering
\caption{The leptonic models parameters for SNRs.}
\label{tab-2}
\vspace{2mm}
\begin{tabular}{|c|c|c|c|c|c|c|c|}
  \hline\hline
    $  $& d    &        n        & $\alpha1$ & $\alpha2$ & E$_{br}$  &  W$_e$     &Ref   \\
    $  $&(kpc) & (cm$^{-3}$)     &           &           & (GeV)     & (erg)   &        \\

  \hline
  Tycho        &  3.0  &  10  & 2.03  & 2.69  & 0.418 & 2.92$\times$10$^{48}$  & 9,10 \\

  \hline
  Cassiopeia~A &  3.4  &  10  & 1.03  & 2.81  &  6.38 & 6.18$\times$10$^{48}$  & 11     \\

  \hline
  IC~443       &  2.0  &  100 & 1.92  & 3.00  & 12.37 & 3.63$\times$10$^{48}$  & 12     \\

  \hline
   W51C        &  6.0  &  100 & 1.68  & 2.77  & 4.49  & 2.07$\times$10$^{49}$  & 13     \\

  \hline\hline
 \end{tabular}
\vspace*{0.5cm}

{
 Notes: W$_e$ is the total energy of electrons.\\

References: {(9) \citet{Giordano2012}; (10) \citet{Zhang2013}; (11) \citet{Abdo2010b}; (12) \citet{Yuan2012}; (13) \citet{Abdo2009b};
}}
\end{table*}

\begin{table*}
\centering
\caption{The hadronic models parameters for SNRs.}
\label{tab-3}
\vspace{2mm}
\begin{tabular}{|c|c|c|c|c|c|c|}
  \hline\hline
    $  $& d    &        n        & $\alpha1$ & $\alpha2$ & E$_{br}$  &  W$_p$  \\
    $  $&(kpc) & (cm$^{-3}$)     &           &           & (GeV)     & (erg) \\

  \hline
  Tycho        &  3.0  &  10  & 1.21  & 2.36  & 1.12   & 1.07$\times$10$^{49}$   \\

  \hline
  Cassiopeia A &  3.4  &  10  & 1.22  & 2.41  & 21.72  & 5.82$\times$10$^{49}$    \\

  \hline\hline
 \end{tabular}
\vspace*{0.5cm}

{
Notes: W$_p$ is the total energy of protons.\\
}
\end{table*}

\subsection{Case studies}

In this subsection we choose several GeV-TeV SNRs as examples to explore
in more details the spectral measurements of their very high energy
$\gamma$-ray emissions by LHAASO, and discuss the physical implications
of these new measurements. Different from that in \S~3.2 whose purpose
is a rough estimate of a big sample, we adopt a physically motivated model,
either leptonic or hadronic, to characterize the $\gamma$-ray emission
of these sources. In order to better fit the observed $\gamma$-ray data,
we assume an exponentially cutoff broken power-law form of the injection
spectrum of accelerated particles
\begin{equation}
  F(E)\propto\exp(-E/E_c)\times\left\{\begin{array}{l}
          E^{-\alpha_1} , E < E_{\rm br} \\
          E^{-\alpha_2} , E \geq E_{\rm br}
  \end{array}
\right.,
\end{equation}
where $E_{\rm br}$ and $E_c$ are the break energy and cutoff energy,
$\alpha_1$ and $\alpha_2$ are the spectral indices below and above
$E_{\rm br}$ respectively.
Based on the current experiment data, both leptonic and hadronic scenarios may still remain
valid if the $\gamma$-ray spectrum has a cutoff between 10 TeV and 100 TeV \citep{Slane2014, Zirakashvili2014, Abdo2010b}.
Because of the shorter cooling time of electrons, the $\gamma$-ray emitting electrons can be hardly accelerated up to 100 TeV.
While there is no strong limit on hadronic acceleration based on the theory or current experiment. Thus, in this paper,
we assume $E_c^e\approx50$ TeV \citep{Yuan2012} in the leptonic model, and $E_c^p=\infty$ in the hadronic model which means that they can be
accelerated to energy above the knee.

\subsubsection{Young SNRs: Tycho and Cassioperia~A}

The relatively young SNRs are believed to be the major sites to accelerate CRs up to 10$^{15}$ eV, and the detection and identification of hadronic $\gamma$-rays directly from young SNRs would be a straightforward test of acceleration of protons and nuclei.

Tycho, which appeared in 1572 \citep{Hanbury1953}, has been observed from radio to TeV $\gamma$-ray band \citep{Dickel1991,Hwang2002,Bamba2005,Stroman2009,Katsuda2010,Acciari2011}.
At the GeV range, Fermi-LAT reported a 5$\sigma$ detection of GeV $\gamma$-rays emission from Tycho, which can be described by a power-law with a photon index 2.3$\pm$0.2 \citep{Giordano2012}.
At the TeV range, VERITAS observed that the total flux of Tycho above 1 TeV is $\sim$ 0.9$\%$ of
Crab Nebula and the spectrum index between 1 TeV and 10 TeV is about 1.95$\pm$0.51 in 2011. But in 2015,
the spectrum index is 2.92$\pm$0.46 \citep{Acciari2011,Park2015}.
If the spectral index is about 2 up to 10 TeV as the VERITAS reported in 2011,
it implies that the corresponding spectrum of primary protons continues  without a significant steepening or a cutoff to at least several hundred TeV \citep{Kelner2006,Aharonian2013}.
Due to the large uncertainties of the data sets of Fermi and VERITAS, the energy spectrum from GeV to TeV can be described by a broad range of function which is not enough to explain the high energy $\gamma$-ray emission.

Cassioperia~A (Cas~A), appeared in the sky in 1680, is the youngest of the historical Galactic SNRs \citep{Abdo2010b,Acciari2010}.
 It is one of the best studied objects with both thermal and non-thermal broad-band emission ranging from radio wavelengths to TeV $\gamma$-rays \citep{Aharonian2001,Albert2007b,Acciari2010,Yuan2013}.
TeV $\gamma$-ray observations revealed a rather modest $\gamma$-ray flux, compared to the synchrotron radio through X-ray emission, which further strengthens the argument for a rather high magnetic field. In the GeV range, Fermi-LAT observation suggests that leptonic model can not fit the turnover well at low energy because the bremsstrahlung component that is dominant over IC below 1 GeV has a steep spectrum, and hadronic emission describing the $\gamma$-ray spectrum by a broken power-law is preferred. However, because the observed TeV $\gamma$-ray fluxes have large statistical uncertainties, it can not be judged yet whether the TeV $\gamma$-rays are generated by interactions of accelerated protons and nuclei with the ambient gas or by electrons through bremsstrahlung and inverse Compton scattering. And the maximum energy of the observed TeV $\gamma$-ray is only several TeV, the question whether Cas~A accelerates particles to PeV energy is still open.

At the LHAASO site, the effective observation time is $6.2$ hours per day for Tycho and $6.8$ hours per day for Cas~A with zenith angle less than $45^{\circ}$.
Tycho culminates with a zenith angle of $34^{\circ}$ and Cas~A culminates with a zenith angle of $29^{\circ}$.
The expected spectrum of Cas~A from 0.3 TeV to 1 PeV is shown in Fig. 3, and the parameters for the leptonic and hadronic model are listed in Tab. 2 and Tab. 3 respectively.
According to the expectation results from Fig. 3, we can see that from 300 GeV to 500 TeV, the statistic error of data obtained by LHAASO will be less than 10$\%$.
Due to the Klein-Nishina effect, the spectral dominated by electrons is much softer than the hadronic acceleration above 10 TeV, and the expected result of LHAASO with a low statistic error can give an reasonable explanation on the high energy range. This estimation would be just sufficient to confirm whether the historical SNRs are PeVatron or not, and give the final judgement for the acceleration models.

\begin{figure*} \centering
\includegraphics[width=3.3in,height=3in]{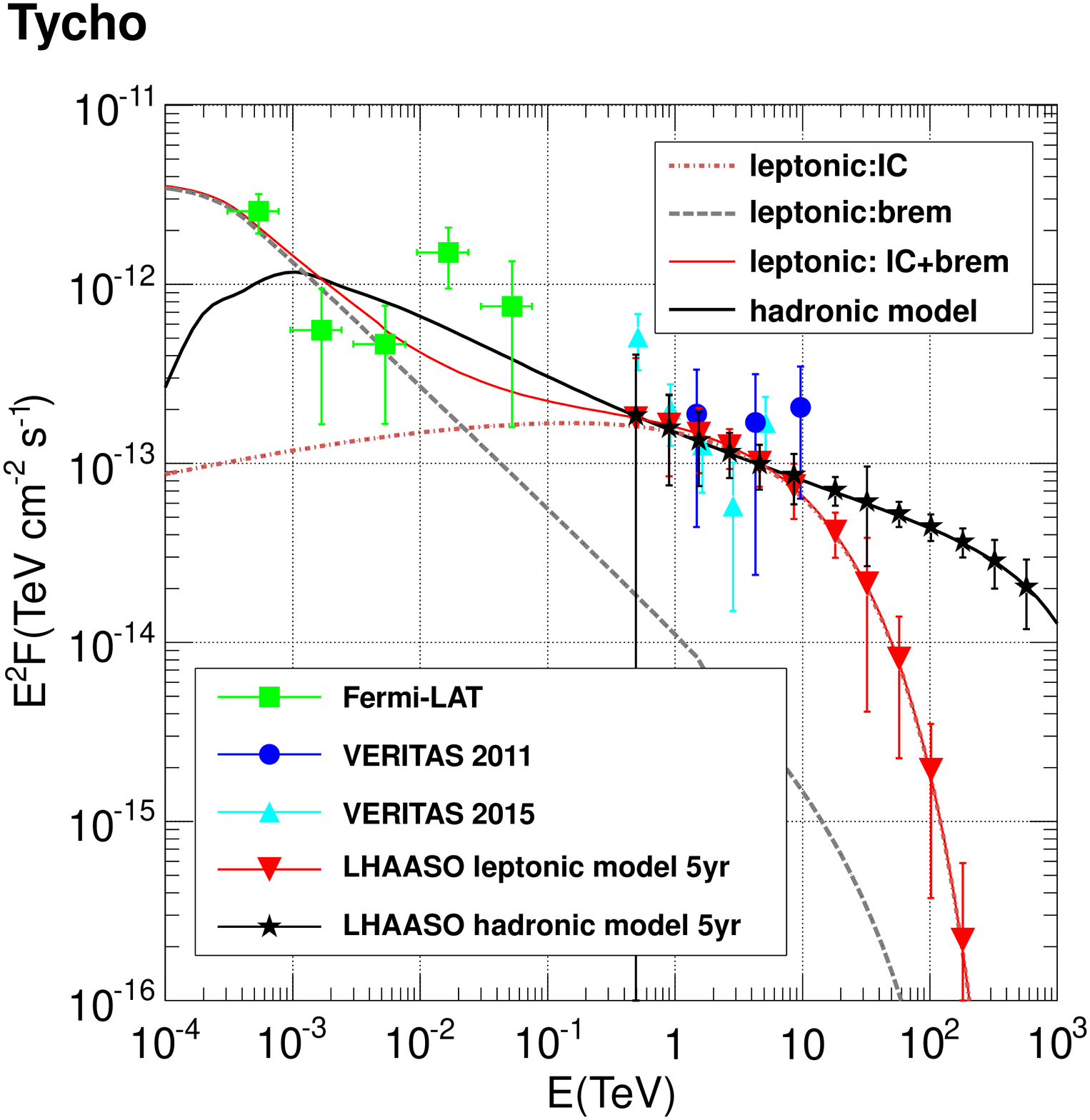}
\includegraphics[width=3.3in,height=3in]{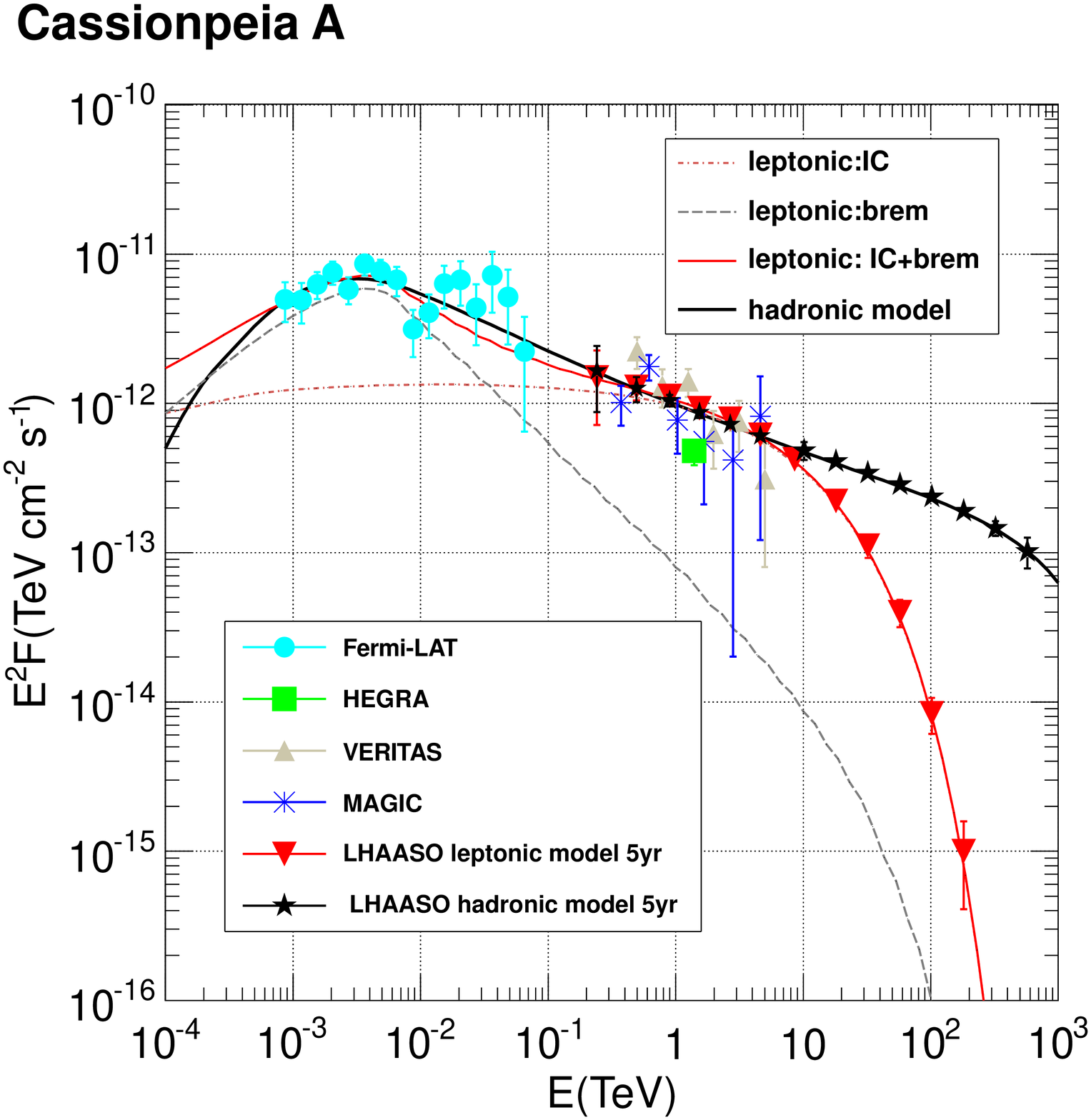}
\caption{Expectation of the LHAASO project on the historical SNR spectrum. (A color version of this figure is available in the online journal.}
\label{fig3}
\vspace*{0.5cm}
\end{figure*}

\begin{figure*} \centering
\includegraphics[width=3.3in,height=3in]{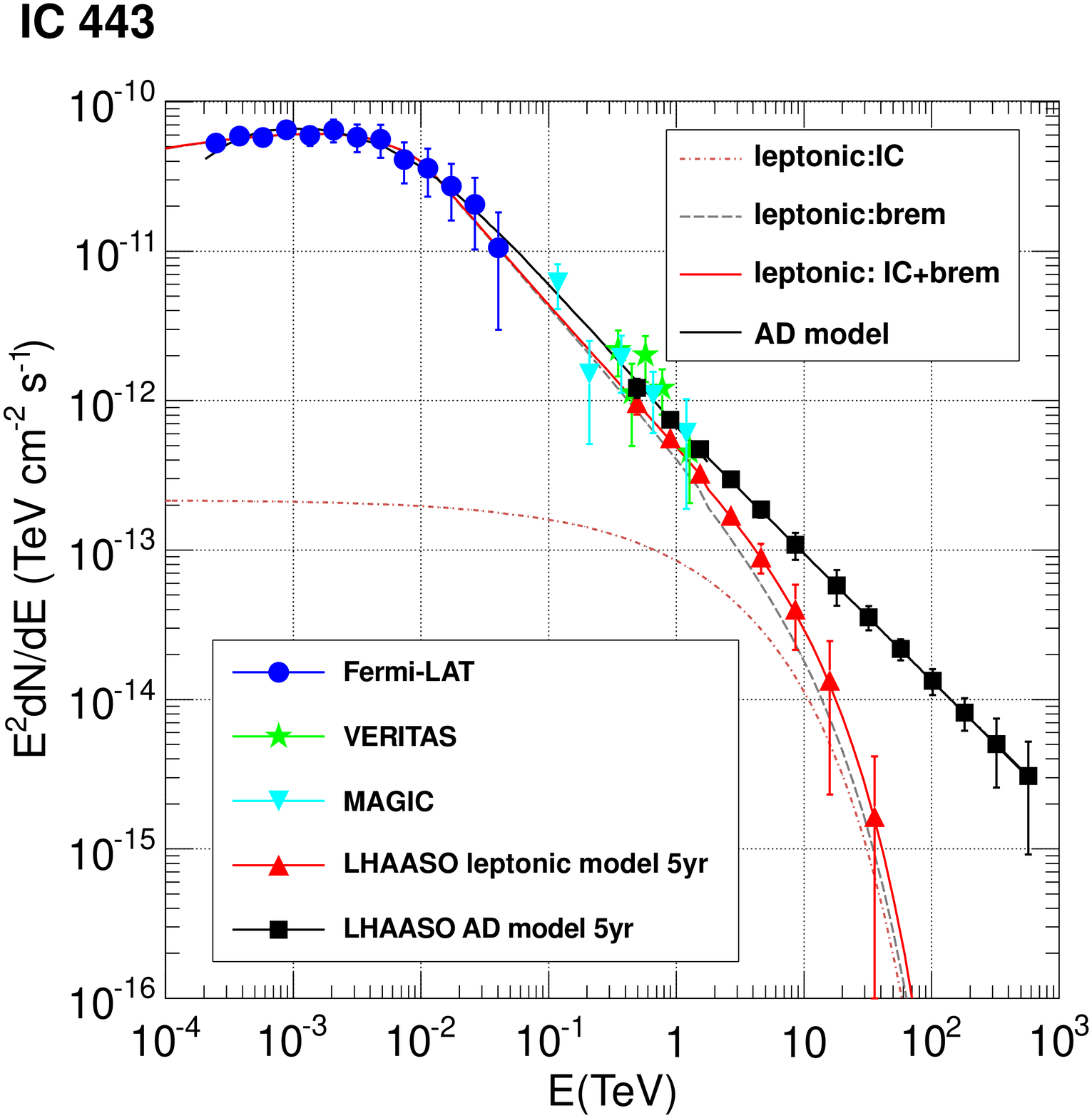}
\includegraphics[width=3.3in,height=3in]{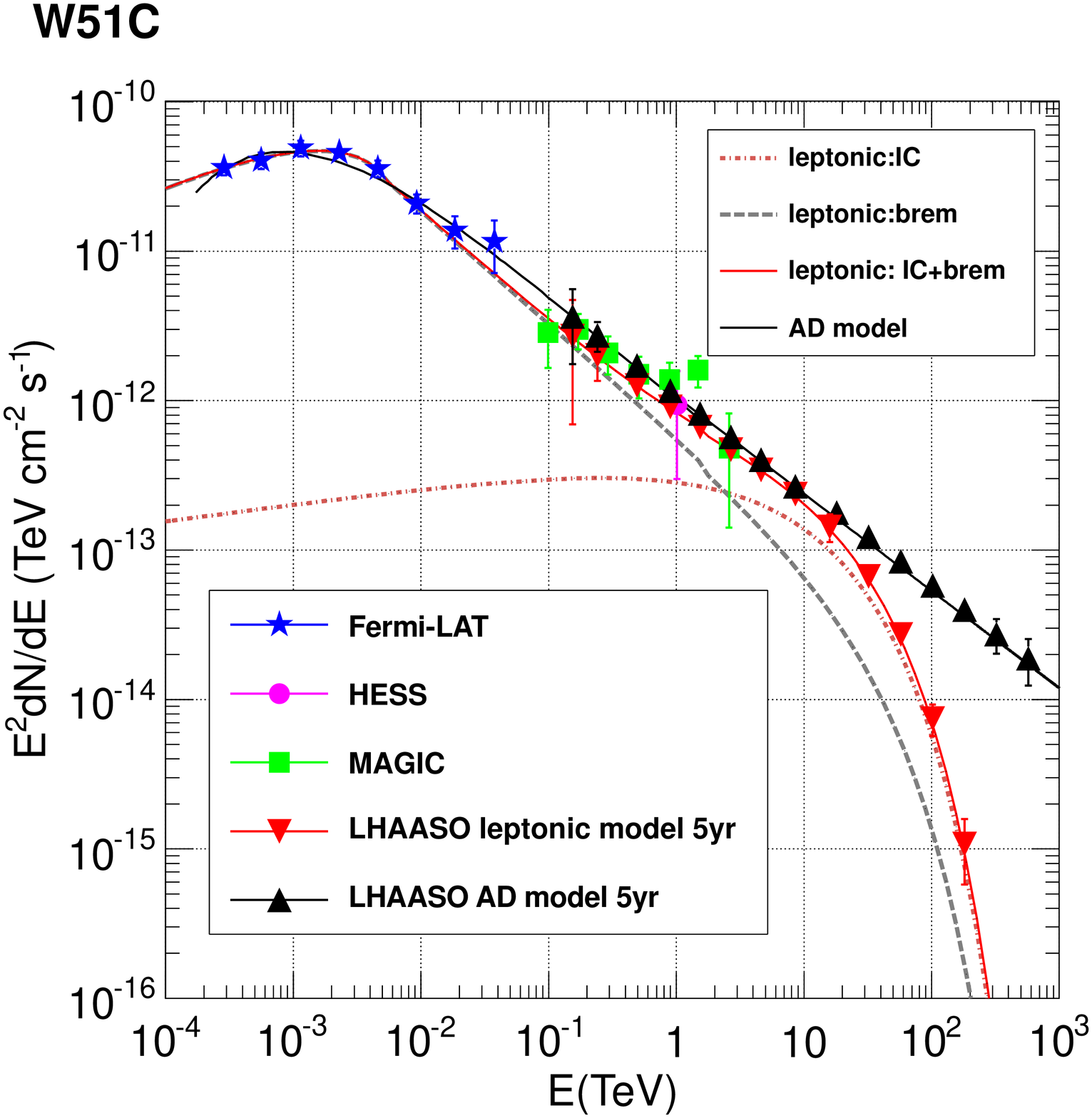}
\caption{Expectation of the LHAASO project on SNRs interaction with molecular clouds spectrum. (A color version of this figure is available in the online journal.}
\label{fig4}
\vspace*{0.5cm}
\end{figure*}

\subsubsection{SNRs interacting with molecular clouds}

The massive molecular clouds located in the vicinity of SNRs, provide dense targets for hadronic interactions, and thus dramatically increase the chances of tracing the run-away protons via the secondary $\gamma$-rays. The location of molecular clouds close to SNRs could be accidental, but in general there is a deep link between SNRs and molecular clouds \citep{Jiang2010}, especially in the star-forming regions \citep{Aharonian1996}.
SNRs interacting with molecular clouds are expected to strengthen $\pi^{0}$-decay emission and then could provide direct evidence of cosmic nuclei being accelerated at supernova shocks \citep{Aharonian1994}.

SNR IC~443 possesses strong molecular line emission regions that makes IC443 a case for an SNR interacting with molecular clouds. The X-ray emission of IC~443 is primarily thermal and peaked towards the interior of the northeast shell, indicating that IC~443 is a SNR with mix-morphology.
Fermi \citep{Abdo2010a} in the GeV region and VERITAS \citep{Acciari2009a}, and MAGIC \citep{Albert2007a} in the TeV region up to 1 TeV detected the $\gamma$-ray spectrum of IC~443, but at higher energy range, there is not yet observation, which is very important for determination on $\gamma$-ray emission mechanism.

W51C (G49.2-0.7) also interacts with the molecular clouds. The W51 region was strong studied as it is known to host several objects. It contains three main components: two star-forming regions W51A and W51B surrounded by very giant molecular cloud, and SNR W51C. W51C is a radio-bright SNR at a distance of ~6 kpc from Earth with an estimated age of $\sim$ 3 $\times$ 10$^{4}$ yr \citep{Koo1995}. W51C is visible in X-rays showing both a shell type and center filled morphology. Shocked atomic and molecular gases have been observed, providing direct evidence on the interaction of W51C shock with a large molecular cloud \citep{Reichardt2011,Aleksic2012}. The GeV spectral result provided by Fermi indicates that leptonic model is difficult to explain $\gamma$-rays production and the most reasonable explanation is that hadronic interaction took place at the shocked shell of W51C emits GeV $\gamma$-rays \citep{Abdo2009b}.
Moreover, MAGIC and H.E.S.S. also indicates the $\gamma$-ray emission from W51C tends to be dominated by $\pi^{0}$-decay up to several TeV \citep{Fiasson2009,Reichardt2011,Aleksic2012}. But this still has uncertainties for the acceleration mechanism above 10 TeV.

Besides of the leptonic model we described above, the accumulative diffusion model is also performed (in this paper, we call it AD model for short) \citep{Li2010,Li2012,Zhang2016}. The AD model considers that the energetic protons colliding with the given molecular clouds are a collection of the diffusive protons escaping from the shock front as the SNR expands. A small distance between the SNR and molecular clouds is allowed \citep{Li2010}. The distribution of escaping protons is assumed to be a power-law function, and the spectrum of the diffusive protons for any point near the SNR at any time is obtained. In the treatment of the dynamical evolution of a SNR, the Sedov-Taylor law is used for the adiabatic phase and the radiative phase. Besides, the finite-volume of the cloud in the vicinity of SNRs is considered. The $\gamma$-rays emitted from the secondary leptons are negligible compared with the dominant contribution of the protons themselves \citep{Gabici2007, Gabici2009}.

The parameters for leptonic model are listed in Tab. 2, and the parameters for AD model are given in the published paper \citep{Li2012}.
At the LHAASO site, the effective observation time is $6.53$ hours per day for IC~443 and $6.0$ hours per day for W51C with zenith angle less than $45^{\circ}$. IC~443 culminates with a zenith angle of $8^{\circ}$ and W51C culminates with a zenith angle of $16^{\circ}$.
The expectation of LHAASO is given in Fig. 4, compared with the measurement of Fermi, MAGIC and VERITAS.
From 300 GeV to 500 TeV, the statistic error of data obtained by LHAASO will be less than 10$\%$. The discrepancy between the expectations from the two models will reach more than 5 sigma above 20 TeV.
It indicates that LHAASO will make a great contribution on the accelerate measurement in the TeV range, providing the final judgement on leptonic or hadronic acceleration.

\subsection{Special case studies}

Most of unidentified TeV sources are extended $\gamma$-ray sources with SNRs, PWNs or PSRs in their region.
The estimation of their significant should include the source extension during the calculation of background events.
To model the source extension, a symmetric two-dimensional Gaussian shape is used for all extended sources in this work. The source extensions $\sigma_{ext}$ for each source is shown in Tab. 4.



\begin{table*}
\centering
\caption{Table of a selection of UNID TeV $\gamma$-ray sourcs and a supperbubble (Cygnus Cocoon) shown with the expected significance above 10 TeV using five years Monte Carlo simulation datas of LHAASO, assuming the sources are under the specific hypotheses for the energy spectrum. (``-" means the significance is less than 5$\sigma$.) }
\label{tab-4}
\vspace{1mm}
\begin{tabular}{ccccccccc}
  \hline\hline
  Name        & R.A.         & Dec.        & $\alpha$     & $J_0$                                 &   $\sigma_{ext}$   & $\sigma$                &$\sigma'$                & Ref.\\
              &              &             &              &     (TeV$^{-1}$cm$^{-2}$s$^{-1}$)     &   ($^{\circ}$)     & ($E_{\rm cut}=100$ TeV) & ($E_{\rm cut}=30$ TeV)  & \\
  \hline
  MAGIC J0223+403       & 02$^h$23$^m$12$^s$       &  +43$^{\circ}$00$^{'}$  & 3.1$\pm$0.31   & 4.07$\times$10$^{-13}$  & 0    &   -     &  -       & 14  \\
  \hline
  HESS J1832-093        & 18$^h$32$^m$50$^s$       &  -09$^{\circ}$22$^{'}$  & 2.6$\pm$0.3    &  4.8$\times$10$^{-13}$  & 0    &  15.27  &  5.15    &15   \\
  \hline
  HESS J1834-087        & 18$^h$34$^m$45.6$^s$     &  -08$^{\circ}$45$^{'}$  & 2.5$\pm$0.2    &  3.7$\times$10$^{-12}$  & 0.1  & 21.73   &   7.15   & 16  \\
  \hline
  HESS J1841-055        & 18$^h$40$^m$55$^s$       &  -05$^{\circ}$33$^{'}$  & 2.32$\pm$0.23  &  3.76$\times$10$^{-11}$ & 0.4  & 699.27  &  234.89  & 17  \\
  \hline
  HESS J1857+026        & 18$^h$57$^m$11$^s$       &  +02$^{\circ}$40$^{'}$  & 2.16$\pm$0.07  & 5.37$\times$10$^{-12}$  & 0.17 & 205.63  & 76.95    & 18  \\
  \hline
  HESS J1858+020        & 18$^h$58$^m$20$^s$       &  +02$^{\circ}$05$^{'}$  & 2.17$\pm$0.12  & 6.0$\times$10$^{-13}$   & 0.08 & 20.10   & 7.45     & 19  \\
  \hline
  MGRO J1908+06         & 19$^h$07$^m$54$^s$       &  +06$^{\circ}$16$^{'}$  & 2.54$\pm$0.36  & 2.06$\times$10$^{-11}$  & 0.49 & 220.80  & 97.13    & 20  \\
  \hline
  VER J2016+371         & 20$^h$16$^m$02$^s$       &  +37$^{\circ}$11$^{'}$  & 2.3$\pm$0.3    & 3.1$\times$10$^{-13}$   & 0    &  9.76   &  5.00    &21   \\
  \hline
  VER J2019+368         & 20$^h$19$^m$25$^s$       &  +36$^{\circ}$48$^{'}$  & 1.75$\pm$0.08  & 1.35$\times$10$^{-12}$  & 2.0  &  58.36  & 22.40    &22   \\
  \hline
  VER J2019+407         & 20$^h$20$^m$04.8$^s$     &  +40$^{\circ}$45$^{'}$  & 2.37$\pm$0.14  & 1.5$\times$10$^{-12}$   & 0.23 &  38.74  & 17.89    & 23  \\
  \hline
  ARGO J2031+4157       & 20$^h$31$^m$12$^s$       &  +42$^{\circ}$30$^{'}$  & 2.6$\pm$0.3    & 3.05$\times$10$^{-12}$  & 2.0  & 10.86   & 5.3     & 24  \\

  \hline\hline
 \end{tabular}
\vspace*{0.5cm}

{ Notes: Columns from left to right are: source name, R.A., Declination, spectral index, flux normalization, cutoff energy,
expected significance by LHAASO (for $E_{\rm cut}=100$ TeV if the cutoff
energy has not been measured), expected significance by LHAASO for
$E_{\rm cut}=10$ TeV, and the references of the measurements.\\
References: {
    (14) \citet{Aliu2009}; (15) \citet{Abramowski2015}; (16) \citet{Albert2006}; (17) \citet{Bartoli2013}; (18) \citet{Aleksic2014}; (19) \citet{Aharonian2008a}; (20) \citet{Bartoli2012}; (21) \citet{Aliu2014}; (22)\citet{Aliu2014}; (23) \citet{Aliu2013}; (24) \citet{Bartoli2014};

 }}
 \end{table*}

 \begin{figure}
 \centering
\includegraphics[width=0.5\textwidth]{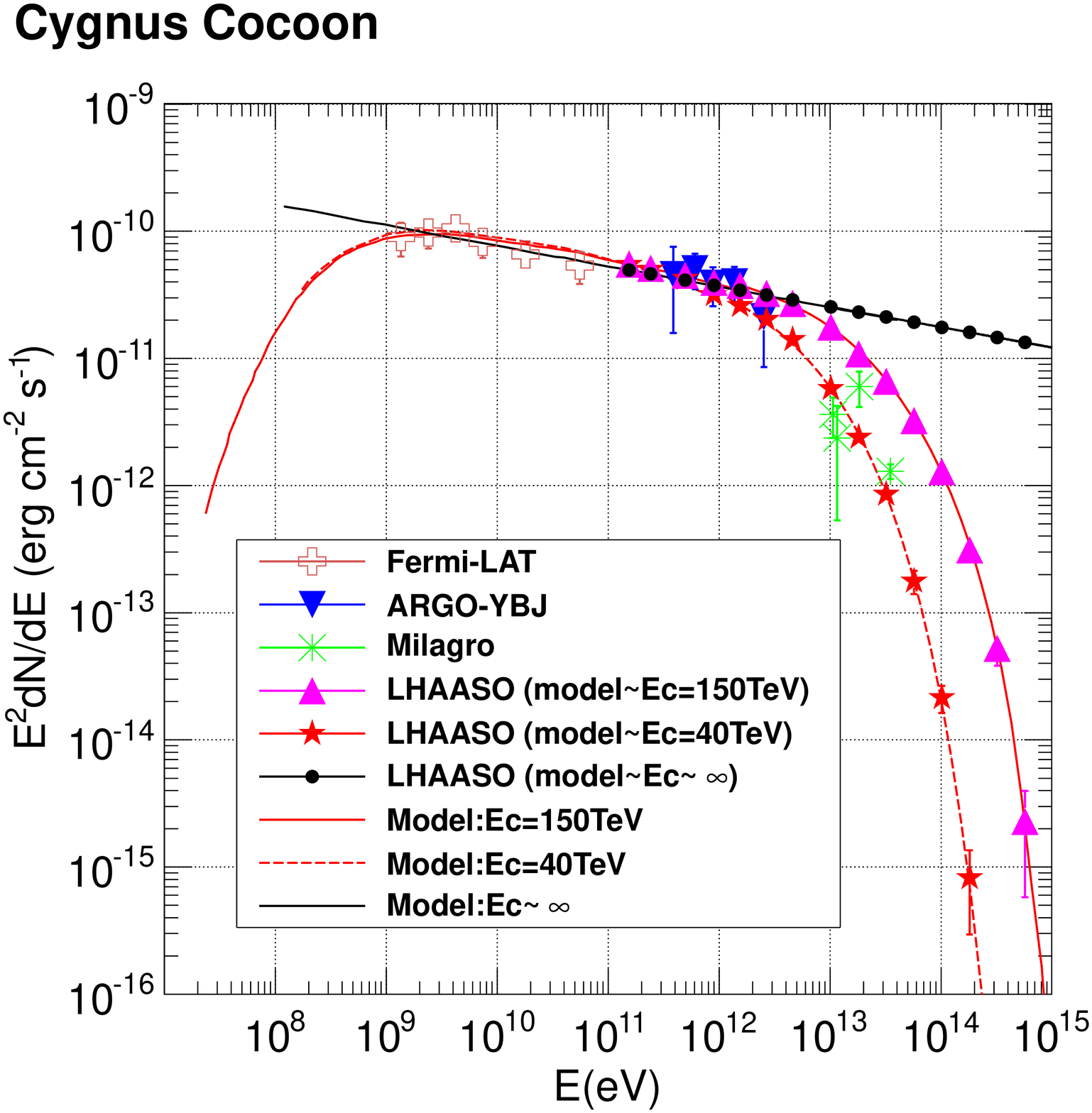}
  \caption{Expectation of the LHAASO project on Cygnus Cocoon by using one year MC data, compared with the measurement of Fermi-LAT(Ackermann et al. 2011), ARGO-YBJ(Bartoli et al. 2014), Milagro(Abdo et al. 2007a; 2007b; 2009a). (A color version of this figure is available in the online journal.)\label{fig5}}
  \label{fig5}
\vspace*{0.5cm}
 \end{figure}

We choose ARGO J2031+4157/Cygnus Cocoon as example to explore in more detail the spectral measurements at high energy range. The Cygnus region of the Galactic plane, discovered by Fermi-LAT at GeV energies in the Cygnus superbubble, is the famous region in the northern sky for the complex features observed in radio, infrared, X-rays, and $\gamma$-rays. It contains a high density interstellar medium, rich in potential cosmic ray acceleration sites such as Wolf-Rayet stars, OB associations, and SNRs. This region is home of a number of GeV $\gamma$-ray sources detected by Fermi-LAT \citep{Nolan2012} and several noteworthy TeV $\gamma$-ray sources detected by Milagro and ARGO-YBJ in the past decade. The Cygnus Cocoon, located in the star-forming region of Cygnus X, is interpreted as a cocoon of freshly accelerated cosmic rays related to the Cygnus super-bubble.
The extended TeV $\gamma$-ray source ARGO J2031+4157 (or MGRO J2031+41) is positionally consistent
with the Cygnus Cocoon and another TeV source MGRO J2019+37 is a mysterious source only being detected by Milagro \citep{Abdo2007a} above 20 TeV and VERITAS \citep{Aliu2014} above 1 TeV. The reason for the hard SED from such a spatially extended region is totally unknown. The discovery of this kind of sources and the more detailed multi-wavelength spectroscopic investigations can be an efficient way to explain the radiation mechanism of them.

At the LHAASO site, Cygnus Cocoon culminates with a zenith angle of $12^{\circ}$ and is observable for 7.2 hours per day.
Fig. 5 shows all the spectral measurements by Fermi-LAT \citep{Ackermann2011}, ARGO-YBJ \citep{Bartoli2014}, Milagro \citep{Abdo2007a, Abdo2007b, Abdo2009a}, and the expectation results with LHAASO.
The hadronic emission model with energy cuts off 150 TeV is the maximum allowed by the ARGO-YBJ upper limit. Taking Milagro data into account, the cutoff energy would be around 40 TeV. One year observation of LHAASO will be sufficient to give a judgement on the different cutoff energy models from 300 GeV to several hundreds TeV. 
Besides, if the spectrum has no cutoff energy shown in Fig. 5 (black dots)
, LHAASO will give an accurate measurement up to PeV. It will provide important information for investigating the particle acceleration within the super-bubble.


\section{Conclusions}

With a sensitivity of 10 mili-Crab at 50 TeV, LHAASO will launch the most accurate detection of $\gamma$-ray spectrum of the SNRs in the energy region above 30 TeV. In the field of view of LHAASO, 8 identified GeV-TeV or TeV SNRs observed by previous experiments will be detected by LHAASO with precise spectrum unprecedented extended up to few hundred TeV. Besides, LHAASO has potential to discover more TeV SNRs which has not yet observation at GeV region and can also provide important information on unidentified TeV $\gamma$-ray sources with a high significance. According to the detailed research on the spectrum, different theoretical models can obtain the effective tests by observation of LHAASO. Especially, with the high sensitivity of LHAASO above 30 TeV, it is expected that LHAASO is more sensitive to different leptonic or hadronic models of SNR $\gamma$-ray emission. High statistics and accurate measurement provided by LHAASO will sufficiently offer a final judgement between different SNR theoretical models, to reveal cosmic ray acceleration mechanism in Galaxy.

\section{Acknowledgments}
We acknowledge the supports of the NSFC grants (No.11205165, No.11575203, No.11375052, No.11375210, No.11233001) and CAS (U1532258), the Program for New Century Excellent Talents in University (NCET-13-0342), the Shandong Natural Science Funds for Distinguished Young Scholar (JQ201402) and  973 Program grant 2015CB857100.

\clearpage


\end{document}